\documentclass[smallextended,referee,envcountsect]{svjour3} 

\usepackage{float}                    
\usepackage{booktabs}            

\usepackage{mathtools}

\usepackage{amssymb}
\usepackage{amsmath}

\usepackage{graphicx}
\usepackage{xcolor}      
 
\usepackage[normalem]{ulem}     

\floatstyle{ruled}
\newfloat{algorithm}{tbp}{loa}
\providecommand{\algorithmname}{Algorithm}
\floatname{algorithm}{\protect\algorithmname}

\numberwithin{equation}{section}
\numberwithin{figure}{section}

\newcommand{\OB}{{\mathcal OB}}
\newcommand{\X}{{\mathcal X}}
\newcommand{\R}{\mathbb{R}}

\DeclareMathOperator{\grad}{grad}
\DeclareMathOperator{\Hess}{Hess}

\DeclareMathOperator{\diag}{diag}
\DeclareMathOperator{\trace}{tr}

\smartqed

\journalname{arxiv}

\begin{document}

\title{Computing Laser Beam Paths in Optical Cavities:}

\subtitle{An Approach based on Geometric Newton Method}

\author{Davide Cuccato \and Alessandro Saccon \and  Antonello Ortolan \and Alessandro Beghi}

\institute{Davide Cuccato, Corresponding author \at
		Department of Information Engineering, University of Padova \\
		Via Gradenigo, 6 35100 Padova, Italy \\
		cuccatod@dei.unipd.it 
\and 
 	Alessandro Saccon \at
	Department of Mechanical Engineering, Eindhoven University of Technology\\
	De Rondom 70, 5612 AP, Eindhoven, the Netherlands \\
	a.saccon@tue.nl
\and	
 	Antonello Ortolan \at
 	INFN - National Laboratories of Legnaro \\
 	Viale dell'Universit\'a, 2 35020 Legnaro (PD), Italy \\
 	antonello.ortolan@lnl.infn.it 
 \and
	Alessandro Beghi \at
 	Department of Information Engineering, University of Padova \\
 	Via Gradenigo, 6 35100 Padova, Italy \\ 
 	beghi@dei.unipd.it}

\date{Received:date / Accepted: date}

\maketitle

\begin{abstract}
In the last decade, increasing attention has been drawn to high precision optical experiments, which push resolution and accuracy of the measured quantities beyond their current limits. This challenge requires to place optical elements (e.g. mirrors, lenses, etc.) and to steer light beams with sub-nanometer precision. Existing methods for beam direction computing in resonators, e.g. iterative ray tracing or generalized ray transfer matrices, are either computationally expensive or rely on overparametrized models of optical elements. By exploiting Fermat's principle, we develop a novel method to compute the steady-state beam configurations in resonant optical cavities formed by spherical mirrors, as a function of mirror positions and curvature radii. The proposed procedure is based on the geometric Newton method on matrix manifold, a tool with second order convergence rate that relies on a second order model of the cavity optical length. As we avoid coordinates to parametrize the beam position on mirror surfaces, the computation of the second order model does not involve the second derivatives of the parametrization. With the help of numerical tests, we show that the convergence properties of our procedure hold for non-planar polygonal cavities, and we assess the effectiveness of the geometric Newton method in determining their configurations with high degree of accuracy and negligible computational effort.
\end{abstract}



\section{Introduction\label{sec:Introduction}}

In advanced optics applications, it is required to optimize the geometry of optical elements, in particular, when very high performance is sought for. One of such applications regards the design of resonant optical cavities that are essential elements of a wide range of devices and experiments, e.g. in laser physics, angular metrology, atomic clocks stabilization, etc. In this paper we focus on \emph{ring laser gyroscopes}, which are devices used for measuring angular rotations with very high accuracy \cite{GW1,GP1,GP2}. The core element of a ring laser is a three-dimensional resonant optical cavity, formed by $N>2$ mirrors that are placed at the vertices of a polygon. In the cavity, the laser beam travels a closed optical path, defining a polygon of perimeter $p$ and area $\mathbf{a}$ \cite{GP1}. In rotation sensing, $\mathbf{a}$ and $p$ are the most relevant geometric quantities, since their value and stability define the device performance in terms of sensitivity and 
accuracy. 

Generally speaking, increasing the cavity dimensions (i.e. $p$ and $\mathbf{a}$) results in measuring devices with higher sensitivity. In fact, the intrinsic noise limiting the sensitivity of a ring laser is the shot noise, and its magnitude turns out to be inversely proportional to $p$. However, the increase of dimensions negatively affects the ring laser long term stability, since changes of the environmental conditions (e.g. temperature and pressure drifts) during the measurement process induce geometry deformations which result in beam-jittering noise with magnitude almost proportional to $p$ \cite{GW1,GW3}.  

Even if a trade off can eventually be made between the intrinsic and beam-jittering noises, to increase sensitivity and stability of an optical cavity, the latter noise must be reduced as much as possible. To this aim, different approaches can be taken, leading to monolithic or heterolithic designs of the optical cavity. In the monolithic approach, one exploits an ultra low expansion material (e.g. Zerodur or Invar) to form a ``rigid frame'' supporting the mirrors, thus achieving passive stabilization of the cavity geometry by regulating pressure and temperature of the environment. For instance, the four-meter-wide square cavity of ``G'' \cite{GW3} (presently the most sensitive and stable ring laser for geodetic and seismic applications) has a monolithic design. In the heterolithic design, mirrors are fixed to a concrete or granite frame and equipped with handlers to react against changes in their relative positions, thus stiffening the geometry of the apparatus \cite{GW2,GP3}. Geometry can also be optimized to reduce the cavity sensitivity to the beam-jittering noise, e.g. by adjusting the beams path to regular polygonal shapes \cite{GP4,GP5}. The heterolithic design overcomes the limitations due to the maximum size of a machinable monolithic element, and it is therefore chosen for very high performance applications, such as fundamental physics \cite{GP6}, geodesy and geophysics \cite{ROMI}. Clearly, implementation of the active geometry control of a heterolithic optical cavity requires the identification of suitable signals, provided e.g. by some metrological precision system, proportional to mirror displacements, and the derivations of accurate models.

In this paper we derive a geometrical model of an heterolithic ring laser to efficiently calculate the beams configuration as a function of the mirror positions and orientations. This problem has already been addressed in the literature. In particular, generalized ray transfer matrices analysis, based on the optical axis perturbation, has been used in \cite{CINABCD}, whereas  iterative ray tracing methods are used in \cite{ITM}. These approaches are based on overparametrized models of the ring laser mirrors or involve a large number of iterations. To overcome these limitations, we exploit the Fermat's principle and the geometric Newton algorithm on matrix manifold. This tool has second order convergence rate and relies on a second order model of the objective function, that in the problem at hand is the cavity length $p$. In particular, the light path in a square cavity made of spherical mirrors is calculated, starting from the positions of their centers of curvature and the value of their curvature radii. As we avoid the use of coordinates to parametrize the beam position on mirror surfaces, the computation of the second order model does not involve the second derivatives of the parametrization. We show that the convergence properties of our procedure hold for the optical cavities of interest. Finally, we assess the effectiveness of the geometric Newton method in determining their configurations with high degree of accuracy and negligible computational effort.
 
The paper is organized as follows. In Sect. 2 notations and definitions are given. Sect. 3 is devoted to the problem statement and formulation. In Sect. 4 we review the geometric Newton Algorithm, which is then specialized in Sect. 5 to the Oblique Manifold.
In Sect. 6 the application of the proposed algorithm to the square ring laser cavity is presented, and in Sect. 7 some numerical results are presented. Conclusions are drawn in Sect. 8.

\section{Notation and Mathematical Preliminaries}
In this paper we make use of the theory of finite dimensional smooth manifolds and covariant differentiation as presented in \cite{AMS08,MARS02}. The symbols in Tab. \ref{tab:notation} will be used throughout the paper.

\subsection*{\emph{Notation}}
\begin{table}[H]\label{tab:notation}
\noindent \raggedright{}
\begin{tabular*}{15cm}{@{\extracolsep{\fill}}ll}
$\mathcal{E}$ & Euclidean space $\mathcal{E}$.\tabularnewline
\addlinespace
$\mathcal{M},\mathcal{N}\subseteq\mathcal{E}$ & Embedded Submanifolds $\mathcal{M},\mathcal{N}\subseteq\mathcal{E}$
.\tabularnewline
\addlinespace
$x\in\mathcal{M}$ & Element $x$ of the manifold $M$.\tabularnewline
\addlinespace
$f:\mathcal{M}\rightarrow\mathbb{R}$, $\bar{f}:\mathcal{E}\rightarrow\mathbb{R}$ & Real valued functions on $\mathcal{M}$ and $\mathcal{E}$ such that $\bar{f}(x)=f(x)$ for $x \in \mathcal{M}$.\tabularnewline
\addlinespace
$\mathfrak{F}(\mathcal{M})$ & The set of smooth real valued functions on $\mathcal{M}$.\tabularnewline
\addlinespace
$\mathfrak{F}_{x}(\mathcal{M})$ & The set of smooth real valued function defined near $x\in\mathcal{M}$.\tabularnewline
\addlinespace
$T_{x}\mathcal{M}$ & The tangent space to $\mathcal{M}$ at $x\in\mathcal{M}$.\tabularnewline
\addlinespace
$\xi_{x}\in T_{x}\mathcal{M}$ & The tangent vector $\xi_{x}$ to $\mathcal{M}$ at $x$. \tabularnewline
\addlinespace
$\mathfrak{X}_{x}(\mathcal{M})$ & The set of smooth vector fields on $\mathcal{M}$ near $x$. \tabularnewline
\addlinespace
$\xi\in\mathfrak{X}_{x}(\mathcal{M})$ & Smooth vector field $\xi:x\mapsto\xi_{x}$ on $\mathcal{M}$ at $x$.\tabularnewline
\addlinespace
$DF:T_{x}\mathcal{M}\rightarrow T_{F(x)}\mathcal{N}$ & The tangent map of $F:\mathcal{M}\rightarrow\mathcal{N}$ at $x$.\tabularnewline
\addlinespace
$\partial f(x)\in T_{x}\mathcal{E}$ & Euclidean Gradient of $f:\mathcal{E}\rightarrow\mathbb{R}$. \tabularnewline
\addlinespace
$\,\partial^{2}f(x):\, T_{x}\mathcal{E}\rightarrow T_{x}\mathcal{E}$ & Euclidean Hessian of $f$. \tabularnewline
\addlinespace
$\left\langle \cdot,\cdot\right\rangle _{\mathcal{M}}: T_{x}\mathcal{M}\times T_{x}\mathcal{M}\rightarrow\mathbb{R}$ & Riemannian metric on $\mathcal{M}$.\tabularnewline
\addlinespace
$\nabla: T_{x}\mathcal{M}\times\mathfrak{X}_{x}(\mathcal{M})\rightarrow\mathfrak{X}_{x}(\mathcal{M})$ & Riemannian connection on $\mathcal{M}\subset\mathcal{E}$.\tabularnewline
\addlinespace
grad$\, f(x)\in T_{x}\mathcal{M}$ & Riemannian Gradient of $f:\mathcal{M}\rightarrow\mathbb{R}$. \tabularnewline
\addlinespace
Hess$\, f(x): T_{x}\mathcal{M}\rightarrow T_{x}\mathcal{M}$  & Riemannian Hessian of $f$.\tabularnewline
\addlinespace
$\mathbb{S}^{2}$ & Unit Sphere $\left\{ \mathbf{x}\in\mathbb{R}^{3},\ \mathbf{x}^{T}\mathbf{x}=1\right\} $\tabularnewline
\addlinespace
$\wedge$  & Canonical cross product between vectors of $\mathbb{R}^{3}$.\tabularnewline
\addlinespace
$\otimes$  & Kronecker product of two matrices,\tabularnewline
\addlinespace
$\diag:\,\mathbb{R}^{n\times n}\rightarrow\mathbb{R}^{n\times n}$ & 
$[\diag(A)]_{ij} = 
\left\{ 
   \begin{array}{ll}
   A_{ij} &  \text{if} ~ i = j, \\ 
   0       & \text{otherwise}
   \end{array}
\right.
$
\tabularnewline
\addlinespace
\end{tabular*}
\end{table}

The Riemannian connection $\nabla$ is determined by the condition

\begin{align}
D\left\langle \xi_{x},\chi_{x}\right\rangle _{\mathcal{M}}[\eta_{x}]= & \left\langle \left(\nabla_{\eta_{x}}\xi\right)_{x},\chi_{x}\right\rangle _{\mathcal{M}}+\left\langle \xi_{x},\left(\nabla_{\eta_{x}}\chi\right)_{x}\right\rangle _{\mathcal{M}}\ ,\label{eq:RCOND}
\end{align}
\noindent
where $x\in\mathcal{M}$, $\eta_{x}\in T_{x}\mathcal{M}$, and $\xi,\chi\in\mathfrak{X}_{x}(\mathcal{M})$, i.e. it is the unique connection that is compatible with the Riemannian metric $\left\langle \cdot,\cdot\right\rangle _{\mathcal{M}}$ of $\mathcal{M}$. Moreover, if $\mathcal{M}=\mathcal{E}$, we have that the covariant derivative associated to the Riemannian connection of the vector field $\xi$ with respect to $\eta_{x}$ is simply the directional derivative, i.e. $\nabla_{\eta_{x}}\xi{\displaystyle =D\xi[\eta_{x}]}$.

\section{Problem Statement and Formulation}

\begin{figure}
\centering
\includegraphics[width=8cm]{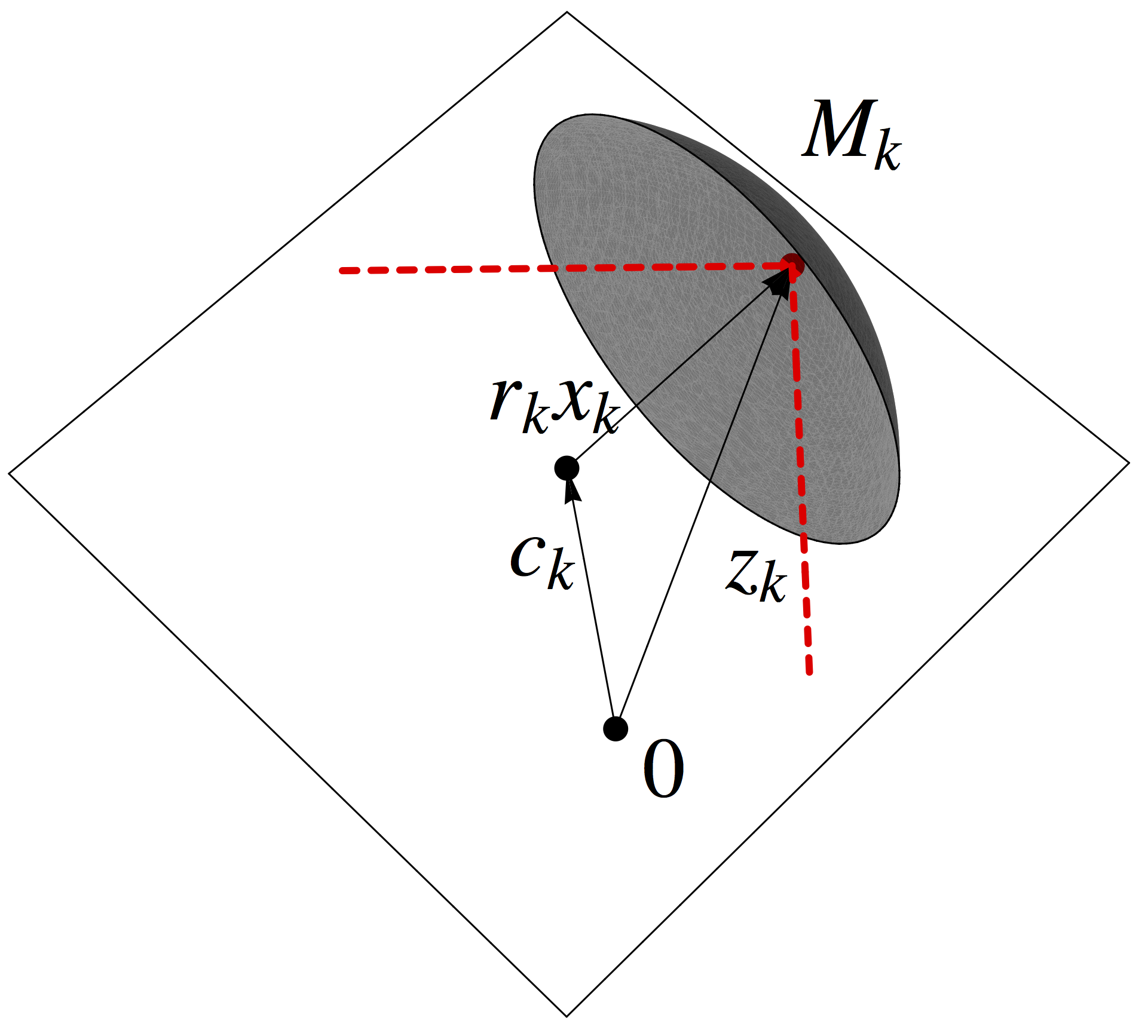}
\protect\caption{\label{fig:1-1} Parametrization of the position of the laser spot on the $k$-th mirror $M_k$. With respect to the inertial frame origin $O$, the position of laser spot is $\mathbf{z}_{k}=r_{k}\mathbf{x}_{k}+\mathbf{c}_{k}$. The mirror $M_{k}$ has radius $r_{k}$. Relative to the mirror center
$\mathbf{c}_{k}$, the position of the laser spot can be parametrized as $r_{k}\mathbf{x}_{k}$, $\mathbf{x}_{k}\in\mathbb{S}^{2}$.} 
\end{figure}

We consider an optical cavity formed by $N$ spherical mirrors. By indicating with $\mathbf{z}_{k} \in \mathbb{R}^3$ the coordinates of the position of the light spot on the \emph{k-th} mirror with respect to the ground frame, the cavity optical length can be computed as  $p=\sum_{k=1}^{N}\parallel\mathbf{z}_{k}-\mathbf{z}_{k+1}\parallel$ and the associated vector area (with magnitude equal to the area of the cavity and direction perpendicular to the cavity plane) as $\mathbf{a}=\frac{1}{2}\sum_{k=1}^{N}\mathbf{z}_{k}\wedge\mathbf{z}_{k+1}$, where we pose $\mathbf{z}_{N+1}\coloneqq\mathbf{z}_{1}$. By applying the formalism of geometric optics, we can model the $k$-th spherical mirror $M_{k}$ as a sphere of center $\mathbf{c}_{k}\in\mathbb{R}^{3}$ and curvature radius $r_{k}\in\mathbb{R}^{+}$. The position of the laser spot on the $k-$th mirror can be expressed as $\mathbf{z}_{k}=\mathbf{c}_{k}+r_{k}\mathbf{x}_{k}$, where $\mathbf{x}_{k}\in\mathbb{S}^{2}$, see  Figure \ref{fig:1-1}.

A \emph{configuration} for the laser beams in the optical cavity is defined as the ordered set of points ${\cal  X }=\left(\mathbf{x}_{1},\dots,\mathbf{x}_{N}\right) \in \underbrace{\mathbb{S}^2 \times \dots \times \mathbb{S}^2}_{N \text{ times}}$, which describe the positions of laser light spots on each of the $N$ mirrors. In addition, we define the matrix of centers ${\cal C}=\left(\mathbf{c}_{1},\dots,\mathbf{c}_{N}\right)$ and the matrix of curvature radii ${\cal R}=diag\left(r_{1},\dots,r_{N}\right)$. We refer to ${\cal R}$ and ${\cal C}$ as the \emph{parameters} of the optical cavity. Given the cavity parameters ${\cal R }$ and ${\cal C }$ and a configuration ${\cal X }$, the corresponding laser beams path as well as the associated (scalar and vector) fields $p$ and $\mathbf{a}$ (see Section \ref{sec:Introduction}) can be computed. 

Note that not every configuration ${\cal X}$  is admissible as physical solution. In fact, the physical light paths have to be stationary points of $p$ with respect to ${\cal X}$  \cite[Chap. 2]{Fermat}. This fact results from the application of the Hamilton's principle to geometric optics, when we identify the Hamiltonian coordinates with the Euclidean coordinates of the laser spots and the action functional with the light optical path length. This result is known in optics as the Fermat's principle, stating that the admissible paths are the ones that make their length stationary with respect to infinitesimal variations of the paths.

We turn now to the problem of computing the physical beam configuration given the cavity parameters ${\cal R}$ and ${\cal C}$. To this end, we resort to an intrinsic geometric approach that avoids a local parametrization of the configuration manifold, only constraining the laser spots on each mirror to lie on the surface of a sphere. The advantages of this approach are to handle simpler expressions and to avoid the use of multidimensional spherical coordinates and derivatives of the parametrization. 

The optical path length is a function of the cavity configuration ${\cal X}$ 
and of the cavity parameters ${\cal C}$ and ${\cal R}$. The configuration ${\cal X}$ lives in the cartesian product of $N$ unit spheres in $\mathbb{R}^3, \underbrace{\mathbb{S}^2 \times \dots \times \mathbb{S}^2}_{N \text{ times}}$.
It is straightforward to show that this configuration manifold is an embedded submanifold of $\mathbb{R}^{3\times N}$. This manifold is usually referred to as the Oblique Manifold of dimension $2\times N$ \cite{AMS08}. The Oblique Manifold of dimension $n\times m$ is defined as
\begin{equation}
{\displaystyle \mathcal{OB}(n,m)=\left\{ {\cal X}\in\mathbb{R}^{\left(n+1\right)\times m},\ d\left( {\cal X}^{T}{\cal X}\right)=I_{m\times m}\right\} \cong\underbrace{\mathbb{S}^{n}\times\dots\times\mathbb{S}^{n}}_{m}}\ ,
\end{equation}
where the function $d$ is defined in Tab. \ref{tab:notation}, and the tangent space at ${\cal X}$ of $\mathcal{OB}(n,m)$ is
\begin{equation}\label{eq:tsobman}
T_{\cal X}\mathcal{OB}(n,m)=\left\{ Y\in\mathbb{R}^{(n+1)\times m}\;:\ d({\cal X}^{T}Y)=0_{m\times m}\right\} .
\end{equation}

It is worth mentioning that  $\mathcal{OB}(n,m)$ inherits the (Riemannian)
metric of $\mathbb{R}^{n\times m}$, i.e., we can define its Riemannian metric simply as $\left\langle {\cal X,Y} \right\rangle _{\mathcal{OB}(n,m)}=\text{tr}\left({\cal X}^{T}{\cal Y}\right)$. Moreover, each column of ${\cal X}$  in (\ref{eq:tsobman}) is orthogonal to the corresponding
column of ${\cal Y}$, with respect to the (Riemannian) metric of $\mathbb{R}^{n+1},$
$\left\langle \mathbf{x},\mathbf{y}\right\rangle _{\mathbb{R}^{n+1}}=\mathbf{x}^{T}\mathbf{y}$. 
In fact the columns of ${\cal X}$ have the straightforward interpretation of points on $\mathbb{S}^2$, so that each column of ${ \cal Y}$ represents a possible tangent vector to the corresponding column of ${\cal X}$, seen as a point of $\mathbb{S}^2$.  

A \emph{physical configuration} $\hat{\cal X}$ is a cavity configuration that also satisfies Fermat's principle, such that it is a stationary point of the optical cavity length $p=p({\cal X;C,R})$. Formally, the set of physical configurations is given by
\begin{equation}
\left\{ {\cal X} \in\mathcal{OB}(2,N)\;:\; T_{\cal X}\mathcal{OB}(2,N)\ni\text{grad }p({\cal X;C,R})=0\right\} .\label{eq:fp}
\end{equation}
By the Weiestrass theorem the set (\ref{eq:fp}) can contain $2$ or more elements being $p$ a continuous function defined over a compact set $\mathcal{OB}(2,N)$. It is not possible, in general, to find closed form expressions for the elements of the set of physical configurations, therefore we resort to a numerical algorithm. We consider optical cavities with parameters ${\cal C}$ and ${\cal R}$ which slightly differ from the nominal values ${\cal C^*,R^*}$. We assume also that the physical configuration ${\cal X}^*$ of the optical cavity with parameters ${\cal C^*,R^*}$ is known. Then, the physical configuration ${\cal X^*}$ can be used as the initial condition for our algorithm. To efficiently compute a solution with desired accuracy, we propose to use a geometric Netwon's method, which requires the computation of first and second order derivatives of the functions $p: \mathcal{OB}(2,N) \mapsto \mathbb{R}$. Such computation can be carried out in an elegant and efficient way by using advanced tools from differential geometry, involving the curvature and affine connection associated with the Riemannian manifold $\mathcal{OB}(2,N)$. 

\section{Stationary points of a vector field on a Riemannian manifold} 
\label{sec:StationaryPoints}

As introduced in the previous section, our aim is to create an algorithm to find  
physical configurations $\mathcal{X}$ of an optical cavity whose geometry 
is characterized by the mirror center positions $\mathcal{C}$ and curvature radii $\mathcal{R}$. We model the configuration space as the Riemannian manifold $\OB(2,N)$ and we call  $p: \OB(2,N) \rightarrow \R$ the length of the light path associated to a given configuration $\X$. 
In this setting, Fermat's principle is equivalent to finding the {\em stationary} points of the vector field $\grad p$ (a physical configuration $\mathcal X$ is, in general, just a extremal point of the light-path length $p$, and not a minimizer).
This section provides a review on an efficient Newton method that can be used for finding stationary points of a vector field on a Riemannian manifold, {\em in the specific situation where this Riemanninan manifold is embedded into an Euclidean space}. 

This section details the algorithm in its general form, considering an arbitrary smooth Riemannian manifold $\mathcal{M}$ endowed with a Riemannian metric $\left\langle \cdot,\cdot\right\rangle _{\mathcal{M}}$  and a real valued function $f \in \mathfrak{F}(\mathcal{M})$ defined on $\mathcal{M}$ embedded in the Euclidean space $\mathcal{E}$. The specific case $f = p$, $\mathcal{M} = \OB(2,N)$, and $\mathcal{E} = \mathbb{R}^{3\times N}$ will be treated in the next section.

The extensions of Newton method to find stationary points of a function defined on an embedded submanifold of the Euclidean space $\mathcal{E}$ (this trivial manifold corresponding to $\mathbb{R}^{n}$ with the standard norm) has received a substantial amount of attention in the recent years as the exploitation of geometric properties leads to effective, elegant, and efficient algorithms \cite{geom1,geom2,geom3,geom4,geom5}. We follow here the approach as outlined in \cite{AMS08,AMSC14,MANOPT}. 
The typical way of approaching the problem of finding the stationary points of a vector field is to reformulate the root finding problem $\grad p(x) = 0$, $x \in \mathcal{M}$ into the minimization problem $\min h(x) := \left| \grad p(x) \right|^2$, $x \in \mathcal{M}$. Naively, one could then employ Newton method to minimize the newly introduced function $h$ over $\mathcal{M}$ by computing a second order approximation of $h$  based on its gradient and Hessian. This naive approach has the disadvantage of requiring the third order derivatives of the length function $p$ for the computation of the Hessian of $h$. However, an efficient alternative method exists that makes use of both functions $h$ and $p$, just requiring the computation of the gradient and Hessian of $p$. Such algorithm, based on Newton method, exhibits second order convergence rate in a neighborhood of stationary point as long as the Hessian of $p$ in not degenerate.  As shown in \cite{AMS08}, at each iteration a line search algorithm (e.g., Armijo's \cite{ARM66}) is run for the function $h$ along a descent direction that is computed via the gradient and Hessian of $p$. 

\subsection{The algorithm} 

A key ingredient in geometric optimization is the use of a \emph{retraction} \cite[Ch.4]{AMS08}.
A retraction $R_{x}:\,T_{x}\mathcal{M} \mapsto \mathcal{M}$ allows one to use the tangent space $T_x \mathcal{M}$ at a given point $x \in \mathcal{M}$ as a local parametrization of the neighborhood of $x$. Retractions are used in the update step of the geometric algorithm, when a descent direction is turned into the next iterate point. The ideal retraction map is the exponential map associated to the Riemannian connection of $\mathcal{M}$. As computing the exponential map is sometimes prohibitive or time consuming, suitable approximations of the exponential map (that agree up to the second derivative at the origin with the exponential map) can be used, still retaining the second order convergence rate of the algorithm. When the manifold is embedded in an Euclidean space, there is a standard method to obtain a retraction map based on tangent space projection.

The geometric Newton algorithm for finding a stationary point of a vector field on a Riemannian manifold can be described at high level as follows.
\emph{}
\begin{algorithm}[H]
\vspace{1ex}
\emph{Input:} $x_{0}\in\mathcal{M},$ real valued function $f$ on $\mathcal{M}$ \\
\emph{Output:} Sequence of iterates $x_1,x_2,\dots$
\begin{enumerate}
\item ~ [{\em Search direction}~] 
Solve (\ref{eq:NEM}) to get the descent direction $\eta_{x_{k}} \in T_{x_k} M$.
\item ~ [{\em Line Search}~] 
Find $t_{k}$ that approximately solves 
$\arg\underset{\lambda}{\min} \,h\left(R_{x}\left(\lambda\eta_{x_k}\right)\right)$
\item ~ [{\em Update}~] 
Set $x_{k+1} = R_x(t_k \eta_{x_k})$
\end{enumerate}
\emph{\protect\caption{\label{alg:GNA} Geometric Newton method for finding stationary points }
}
\label{alg:NewtonMethodStatPoints}
\end{algorithm}
\noindent
The search direction computation requires the solution of the classical geometric Newton equation (cf. \cite[Ch.6]{AMS08}) 
\begin{equation}\label{eq:NEM}
   \text{Hess}\, f(x_{k})[\eta_{x_{k}}]
   = 
   - \grad f(x_{k}), \quad \eta_{x_{k}}\in T_{x_{k}}\mathcal{M}, 
\end{equation}
where the computation of the Riemannian Hessian requires the Riemannian connection $\nabla$ associated with $M$. 
The search direction computation corresponds to the minimization of the quadratic model 
$\widetilde{f}$ of $f$ centered at the current iterate $x_{k} \in \mathcal{M}$ defined as
\begin{equation}
\widetilde{f}(x_{k},\eta_{x_{k}})=f(x_{k})+\left\langle \text{grad}\, f(x_{k}),\eta_{x_{k}}\right\rangle +\frac{1}{2}\left\langle \text{Hess}\, f(x_{k})[\eta_{x_{k}}],\eta_{x_{k}}\right\rangle .
\end{equation}
After the computation of the descent direction $\eta_{x_k} \in T_{x_k} \mathcal{M}$, a line search algorithm is employed, now on the function $h$. An effective line search algorithm is the
Armijo's \cite{armref}. 
At each iterate, the step size $t$ is set to $\alpha\beta^{l}$ (backtracking approach), with $l$ the smallest integer
such that 
\begin{equation}
h\left(R(t \, \eta_{x})\right)\leq h(x)+ \sigma \, t\, Dh(x)[\eta_{x}]\ ,\label{eq:CARM} 
\end{equation}
with $x = x_k$ denoting the current iterate, $\eta_{x} = \eta_{x_k}$ the current descent direction,
and $\alpha > 0$, and $\beta \in (0,1)$, and $\sigma\in(0,1)$ design parameters.
The condition (\ref{eq:CARM}) assure the convergence of the line search
if the function $h(R(t_{k}\eta_{x}))$ to be minimized is sufficiently
smooth (for a proof, see, e.g., \cite[Chapter 4]{AMS08}).

Once the search direction and the step size have been found, the next iterate is computed and the procedure is repeated as long as a stop criterium is not met (norm of $\grad f$ sufficiently small).
The proof of convergence of Algorithm \ref{alg:GNA} to a stationary point of $\grad f$ can be obtained by a simple adaption of the convergence result presented in \cite[Chapter 4]{AMS08} and is described in the Appendix. 

\subsection{The exploitation of the embedding of $\mathcal{M}$ into $\mathcal{E}$} 
\label{subsec:Px}

In the following, we detail how to solve (\ref{eq:NEM}) exploiting 
the fact that $\mathcal{M}$ is embedded in the Euclidean Space $\mathcal{E}$. 
In particular, such a hypothesis allows treating both points of 
$\mathcal{M}$ and vectors of $T\mathcal{M}$ as elements of 
$\mathcal{E} \simeq T \mathcal{E}$ and 
use the trivial Riemannian connection of $\mathcal{E}$ 
to perform covariant differentiation on $\mathcal{M}$ to obtain the Riemannian Hessian of $f$.

Given a point $x\in\mathcal{M}$, the Newton equation requires the 
solution of the linear system \eqref{eq:NEM} to find the descent direction 
$\eta_{x}$. We recall that, on a manifold, the Riemannian gradient and Hessian 
at $x\in\mathcal{M}$ are defined such that 
\begin{align}
\label{eq:RGH1}
\left\langle \text{grad}\, f(x),\xi_{x}\right\rangle _{\mathcal{M}} & =Df(x)[\xi_{x}]  \\
\label{eq:RGH2}
\text{Hess}\, f(x)[\eta_{x}] & =\left(\nabla_{\eta_{x}}\text{grad}\, f\right)_{x} ,
\end{align}
where $Df$ denote the differential of $f$ and $\nabla$ covariant differentiation 
associated to the Riemannian metric of $\mathcal{M}$. 
When $\mathcal{M}$ is an embedding of $\mathcal{E}$, since we can represent
vectors in $T_x\mathcal{E}$ as points in $\mathcal{E}$ ($T_x\mathcal{E} \simeq \mathcal{E}$) and $T_x\mathcal{M} \subseteq T_x\mathcal{E}$, 
we can represent points in $\mathcal{M}$ and $\mathcal{E}$ and tangent vectors in $T_{x}\mathcal{M}$ and $T_{x}\mathcal{E}$ using the same representation \cite[Chapter 3]{AMS08}.

Computation of the Riemannian gradient and Hessian of a function $f$ defined on $\mathcal{M}$ is achieved by computing the standard gradient and Hessian of 
any smooth extension $\bar f$ of $f$ (a smooth function whose restriction on $M$ corresponds to $f$) and employing the tangent space projection detailed below. The gradient and Hessian of $\bar{f}$ in $\mathcal{E}$ will be denoted $\partial\bar{f}$ and $\partial^{2}\bar{f}$ respectively, and called, from now on, Euclidean gradient and Euclidean Hessian of $\bar f$. The interested reader is referred to \cite[Chapter 5]{AMS08} for further details about the proof of the following results.

\noindent {\bf Tangent Space Projection.}  
For any $x \in \mathcal{M}$, the orthogonal projection operator
\begin{align}
P_{x}:T_{x}\mathcal{E} & \rightarrow T_{x}\mathcal{M}\label{eq:PROJ}\\
\xi_{x} & \mapsto P_{x}(\xi_{x})\nonumber 
\end{align}
maps an arbitrary tangent vector $\xi_{x}$ of $T_{x}\mathcal{E}$ into
a vector of $T_{x}\mathcal{M}$ corresponding to
the orthogonal projection of $T_{x}\mathcal{E}$ into its subspace $T_{x}\mathcal{M}$. 
In addition, every tangent vector $\xi_{x}$ in $T_{x}\mathcal{E}$ can be decomposed
as the direct sum $\xi_x = \eta_x + \nu_x$, where $\eta_{x}\in T_{x}\mathcal{M}$
and $\nu_{x}\in (T_x \mathcal{M})^{\perp}$, the normal
space of $\mathcal{M}$ at $x$, i.e. the orthogonal complement of
$T_{x}\mathcal{M}$ in $T_{x}\mathcal{E}$. 
It follows that if $\eta_{x} = P_{x}(\xi_{x})$ then $v_{x} = \xi_{x} - P_{x}(\xi_{x})$ \cite{AMS08}. 

\noindent {\bf Riemannian gradient.}  
The Riemannian gradient $\grad f$ is computed via the Euclidean gradient $\partial\bar{f}$ and
the tangent space projection  \eqref{eq:PROJ} as
\begin{equation}
\grad f(x) = P_{x}\left(\partial\bar{f}(x)\right) , 
\label{eq:RGM}
\end{equation}
for every $x \in \mathcal{M}$. 

\noindent {\bf Riemannian Hessian.}  
Denote with
\begin{align}
  D \grad \bar{f} (x): T_{x}\mathcal{E}  \mapsto T_{x}\mathcal{E}
  \label{eq:dirder}
\end{align}
the tangent map of $\grad f$ defined in \eqref{eq:RGM} seen as the 
function ${\bar P}_x (\partial \bar f(\cdot)): \mathcal{E} \rightarrow \mathcal{E}$, using
the identification $T_x \mathcal{E} \simeq \mathcal{E}$ and where ${\bar P}_x : T_x \mathcal{E} \rightarrow T_x \mathcal{E}$ is a smooth extension of $P_x$ defined for $x \in \mathcal{E}$. 
Then, the Riemannian Hessian of $f$ at $x$ can be computed as 
\begin{align}
  \Hess f(x)[\eta] & =  P_{x} \left(D\,\text{grad}\,\bar{f}(x)[\eta]\right) 
  \label{eq:RCM}
\end{align}
for all $x \in \mathcal{M}$. We recall that, in general, 
$D\,\text{grad}\,\bar{f}(x)\neq D\,\partial\bar{f}(x)=\partial^{2}\bar{f}(x)$. 

It is not straightforward to give a general expression for
$D\,\text{grad}\,\bar{f}$ in (\ref{eq:RGM}). Usually, this
relation is expressed in terms of the Weingarten map, also known
as shape operator\cite{AMSC14}. In this paper,
we limit ourself to provide the specific expression 
for $D\,\grad \bar{f}(x)$ for $\mathcal{M} = \mathcal{OB}(2,N)$.

\subsection{The special case $\mathcal{M} = \mathcal{OB}(2,N)$ and $\mathcal{E} = \mathbb{R}^{3 \times N}$}

We restrict here our analysis to scalar fields defined on $\mathcal{OB}(2,N)$. As the oblique manifold $\mathcal{OB}(2,N)$ 
can be embedded in the Euclidean space $\mathcal{E}$ of dimension $3\times N$,
a point $x\in\mathcal{OB}(2,N)$ and a tangent vector $\xi_{x} \in T_x \mathcal{OB}(2,N)$ 
can both be represented by vectors in $\mathbb{R}^{3\times N}$. 
We detail in the following how to compute the gradient, Hessian, and retraction required in Algorithm \ref{alg:GNA} for the specific case $\mathcal{M} = \mathcal{OB}(2,N)$, relevant to our application.

As we saw in the previous section (in particular, \eqref{eq:RGM} and \eqref{eq:RCM}), the computation of the Riemannian gradient and Hessian requires the computation of 
the orthogonal projection $P_{x}(\xi_{x})$ with $\xi_{x}\in T_{x} \mathcal{M}$ and
the directional derivative $D\,\text{grad}\,\bar{f}(x)$. 
Indicating with $X \in \mathbb{R}^{3\times}$ the (redundant) parameterization of a point $x \in \mathcal{OB}(2,N)$ and 
with $\xi_X \in \mathbb{R}^{3\times}$ the parametrization of a vector $\xi_x \in T_x\mathcal{OB}(2,N)$, these two linear 
operators can be can computed by the expressions provided in Table \ref{tab:CoordRepOB2N}. 
Note in particular that $\partial \bar f( \mathbf{X} )$ and $\partial^2 \bar f( \mathbf{X} ) [\eta_{\mathbf{X}}]$ are 
$3 \times N$ matrices.
The formulas are obtained by straightforward (matrix) differentiation. 
Recall that the function $\diag: \mathbb{R}^{N \times N} \rightarrow \mathbb{R}^{N \times N}$, 
defined in the notation section, simply extracts the diagonal matrix from a given matrix.
Note furthermore that the term $X \diag ( \eta_{X}^T \partial \bar{f} (X) )$ appearing in 
the expression for $D\,\text{grad}\,\bar{f}(X)[\eta_{X}]$ defined in Table~\ref{tab:CoordRepOB2N}
has no influence in the computation of the Hessian as its projection is zero: its computation can therefore be avoided to save computational time. Further details on how these formulas have been obtained can be found in \cite{MANOPT} 

\begin{table}[ht]
\centering
\begin{tabular}{|l|l|}
\hline
\mbox{Abstract terms} & \mbox{Coordinate representation in $\mathbb{R}^{3 \times N}$}\\
\hline 
$x$ & $X \in \mathbb{R}^{3\times N}$\\
\hline 
$\xi_x$ & $\xi_X \in \mathbb{R}^{3 \times N}$\\
\hline 
$P_{x}(\xi_{x})$ 
& 
$ 
  P_{X}(\xi_{X}) :=  
  \xi_{X} - X \diag\left(X^{T}\xi_{X}\right) \
  \in \mathbb{R}^{3 \times N}
$ \\
\hline 
$\begin{array}{c} D\,\text{grad}\,\bar{f}(x)[\eta_{x}] \\ ~ \end{array} $  
&  
\parbox{0.7\textwidth}{
$ 
\begin{array}{rl}
  D\,\text{grad}\,\bar{f}(X)[\eta_{X}] := &   
  P_{X}\left(\partial^{2}\bar{f}(X)[\eta_{X}]\right) - \eta_{X} \diag\left(X^{T}\partial\bar{f}(X)\right) 
  \\ 
  - & X \diag ( \eta_{X}^T \partial \bar{f} (X) )  \in \mathbb{R}^{3\times N}
\end{array}
$
}
\\  
\hline
\end{tabular}
\caption{\label{tab:CoordRepOB2N}
Coordinate representation for  
$P_{x}(\xi_{x})$ and $D\,\text{grad}\,\bar{f}(x)[\eta_{x}]$ 
for the $\mathcal{OB}(2,N)$.}
\end{table}

\noindent{\bf Solution to the geometric Newton equation}. 

The geometric Newton equation \eqref{eq:NEM} is a linear equation with solution  $\eta_{X} \in T_{X}\mathcal{OB}(2,N) 
$. Here we detail how to solve this abstract problem by choosing, for a given 
configuration $X \in {\mathcal{OB}}(2,N) \subset \mathbb{R}^{3\times N}$, a suitable basis for the tangent space $T_{X}\mathcal{OB}(2,N) \subset T_{X}\mathbb{R}^{3\times N}$, 
turning the linear equation into a classical set of linear equations for the coordinate representation of the ``descent'' 
direction $\eta_{X}$. 

\begin{figure}[htb]
\centering
\begin{tabular}{cc}
\includegraphics[width=6cm]{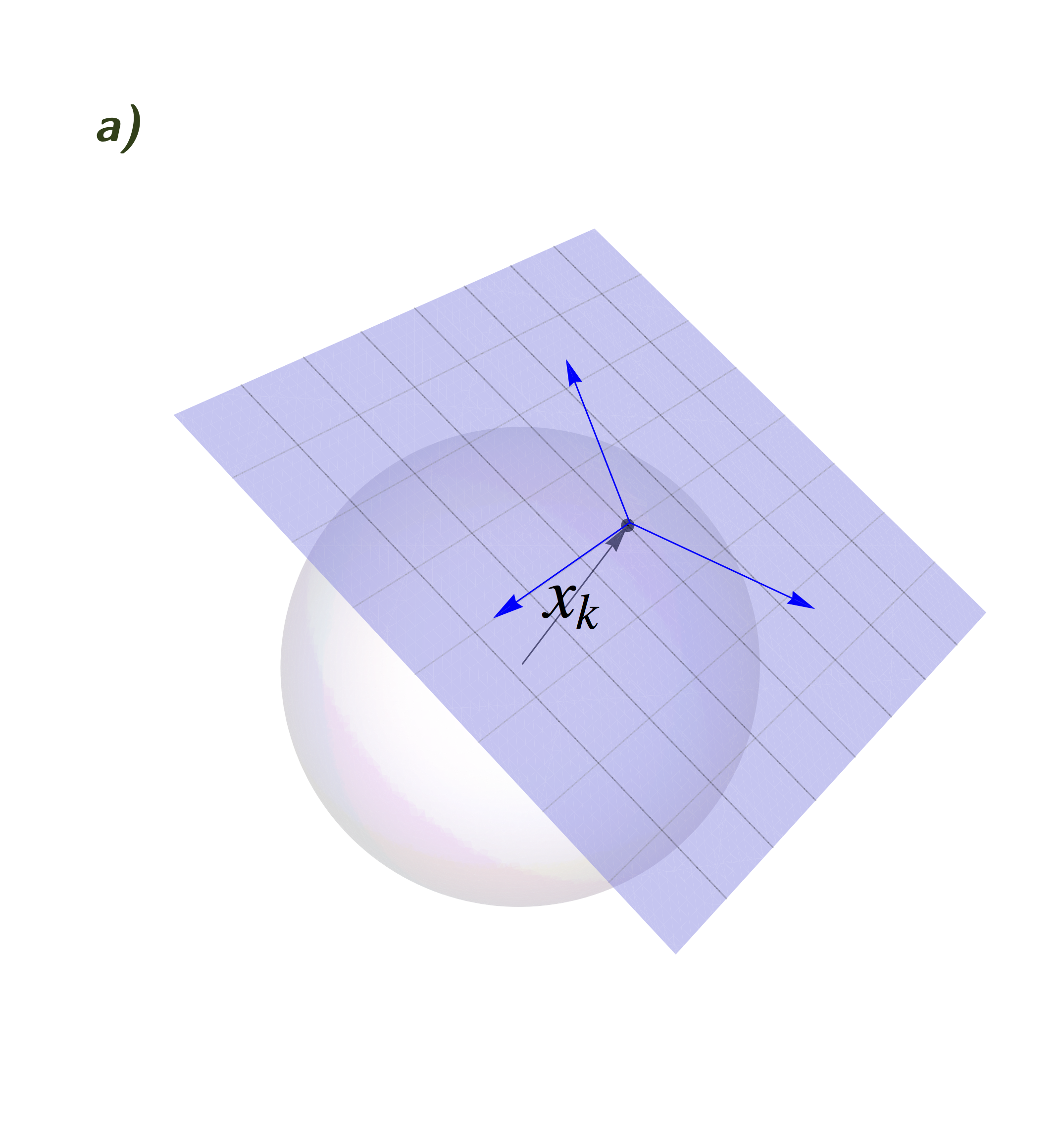} &
\includegraphics[width=6cm]{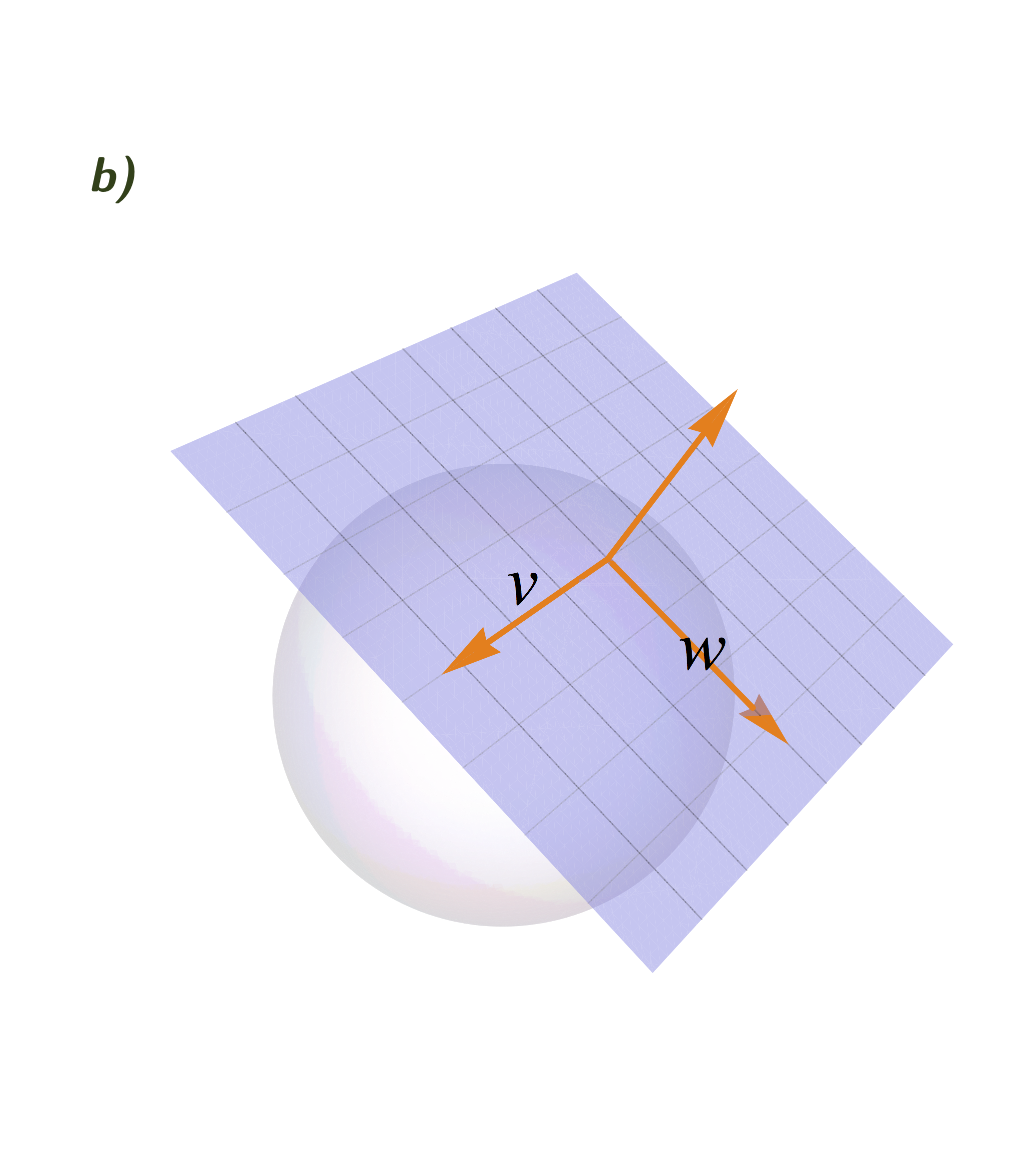} 
\end{tabular}
\caption{
\label{fig:bc} 
a) the point $\mathbf{x}_i$, its tangent space $T_{\mathbf{x}_i}(\mathbb{S}^2)$, and the three row vectors of the matrix which represents $\pi (\mathbf{x}_i)$; b) the tangent space $T_{\mathbf{x}_i}(\mathbb{S}^2)$ and the vectors $v$ and $w$.}  
\end{figure}

When considering $\mathcal{OB}(2,N)$ as embedded in $\mathbb{R}^{3\times N}$, 
a point in $X \in \mathcal{OB}(2,N)$ $\subset \mathbb{R}^{3\times N}$ can be written simply as the matrix
$X = [\mathbf{x}_1,\dots,\mathbf{x}_N]$, with ${\bf x}_i \in  \mathbb{S}^2 \subset \mathbb{R}^3$. 
Recall that  $\mathcal{OB}(2,N)$ and, consequently, $T_{X}\mathcal{OB}(2,N)$ 
for a given $ X \in \mathcal{OB}(2,N)$ have dimension $2N$.
Define $\pi (\mathbf{x}) := I_3 - \mathbf{x} {\mathbf{x}}^T$, $\mathbf{x} \in \mathbb{S}^2$, 
the orthogonal projection of a vector in $\mathbb{R}^3$ into the 2-dimensional subspace of $\mathbb{R}^3$ that is orthogonal to the unit vector $\mathbf{x}$. 
All rows of the matrix $\pi (\mathbf{x})$ are orthogonal to the unit vector $\mathbf{x}$ (for some $\mathbf{x}$, 
one of the row of $\pi (\mathbf{x})$ can be zero). To obtain a basis for $T_{\mathbf{x}} \mathbb{S}^2$, one can pick 
(one of) the row of $\pi(\mathbf{x})$ with maximum norm and identify it as a vector $\mathbf{v}$ that one uses 
as the first element of the basis. 
One then defines  $\mathbf{w}  := \mathbf{v} \times \mathbf{x}$ as the second element of the basis. 
In general, $\mathbf{v}$ and $\mathbf{w}$ are not unit vectors. The selection of $ \mathbf{v} $ and $ \mathbf{w} $ 
as a function of $\mathbf{x}$ cannot be continuous over the entire sphere (this is a consequence of the hairy ball 
theorem), but the lack of continuity is not an issue as one just need {\em a} basis of the tangent space to solve the 
Newton equation $\eqref{eq:NEM}$. A simple visualization of the just mentioned {\em project-and-select} approach is 
given in Figure~\ref{fig:bc}.

As $\mathcal{OB}(2,N)$ is simply the cartesian product of $N$ copies of the unit sphere $\mathbb{S}^2$, 
we can use the same project-and-select approach to construct a basis for $T_{\mathbf{X} }(\mathcal{OB}(2,N)) \subset T_{X}\mathbb{R}^{3\times N}$, $\mathbf{X} = [\mathbf{x} _1, ..., \mathbf{x} _N ]$, by combining $N$ bases obtained with the project-and-select approach for the spaces $T_{\mathbf{x}_i} \mathbb{S}^2$. Formally, denoting $\mathbf{v}_i$ and $\mathbf{w} _j \in \mathbb{R} ^ 3$ 
the basis for $T_{\mathbf{x} _i} \mathbb{S} ^2 $, a basis for $T_{\mathbf{X}} \mathcal{OB}(2,N) \subset \mathbb{R}^{3\times N}$ is the set of $2N$ matrices in $\mathbb{R}^{3\times N}$ defined as $\bigcup_{i=1}^N \Big\{\mathbf{v}_i \otimes \mathbf{e}_i^T, \mathbf{w}_i \otimes \mathbf{e}_i^T \Big\}$, where $\mathbf{e}_i \in \mathbb{R}^N$ denotes the standard $i$-th unit vector of $\mathbb{R}^N$ and $ \otimes $ the Kronecker product. We immediately also 
get a (nonstandard) basis for $T_{\mathbf{X} } R^3 = T_\mathbf{X} \mathcal{OB}(2,N) \oplus T^{\perp}_{\mathbf{X}} \mathcal{OB}(2,N)$ simply adding the orthogonal spaces spanned by $ \mathbf{x} _i $ as 
$\bigcup_{i=1}^N \Big\{\mathbf{x}_i \otimes \mathbf{e}_i^T, \mathbf{v}_i \otimes \mathbf{e}_i^T, \mathbf{w}_i \otimes \mathbf{e}_i^T \Big\}$.

\begin{proposition}
\label{prop:1}
Let $X \in \mathcal{OB}(2,N)$ and $f: \mathcal{OB}(2,N) \rightarrow \mathbb{R}$ with
extension $\bar f: R^{3\times N} \rightarrow \mathbb{R}$ (i.e., $\bar f|_{\mathcal{OB}(2,N)} = f$).
Let $\mathbf{b}_k(X)$, $k = \{1,2, ..., 2N\}$ denote the basis of $T_{X} \mathcal{OB}(2,N)$ 
using the project-and-select approach, then the Newton equation \eqref{eq:NEM} for $f$ with respect to this basis reads
\begin{equation}
  H(X) \mathbf{y} = \mathbf{g}(X)
\end{equation}
with $\mathbf{y} \in \mathbb{R}^{2N}$ is the coordinate representation of $\eta_X$, $\eta_X  = \mathbf{y} _j \, \mathbf{b} _j(X  )$,  and 
the matrix $H( X ) \in \mathbb{R}^{2N \times 2N}$ and $\mathbf{g}(X) \in \mathbb{R}^{2N}$ satisfy 
\begin{align}
  [g(X )]_k ~\mathbf{b}_k(X)        
  & 
  = \grad f( X ) 
  = P_{X} ( \partial \bar f (X )) 
\\
  [H(X )]_{ij} ~ \mathbf{b}_i(X)      
  & 
  = Hess f( X ) [ \mathbf{b} _j(X ) ]
  = P_{X} ( D \grad \bar f (X ) [\mathbf{b}_j(X)] )
\end{align}
where $P_X$ and  $D \grad \bar f (X)$ in $\mathcal{OB}(2,N)$ as given in in Table~\ref{tab:CoordRepOB2N}.
\end{proposition}
The proof of Proposition~\ref{prop:1} is a straightforward exercise based on employing 
the definition of the basis $\mathbf{b}_k(X)$.
Note that the ordering of the basis is completely arbitrary: In our implementation, 
we used 
$\mathbf{v}_1 \otimes \mathbf{e}_1^T$, 
$\mathbf{v}_2 \otimes \mathbf{e}_2^T$,
$...$
$\mathbf{v}_N \otimes \mathbf{e}_N^T$,
$\mathbf{w}_1 \otimes \mathbf{e}_1^T$,
$\mathbf{w}_2 \otimes \mathbf{e}_2^T$,
$...$,
$\mathbf{w}_N \otimes \mathbf{e}_N^T$,
but any other choice would be legitimate.

~\\
\noindent{\bf A retraction for $\mathcal{OB}(2,N)$}.

Despite the fact that for each Riemannian manifold,
the exponential map is a natural retraction, 
different retractions are usually employed to minimize computational costs
without compromising the convergence rate of the zero finding (or minimizing) algorithms. 
For $\mathcal{OB}\left(2,N\right)$, in particular, the map 
\begin{equation}
R(X,\xi_{X})=(X+\xi_{X}) \, \diag \left((X+\xi_{X})^{T}(X+\xi_{X})\right)^{-1/2}\ ,\label{eq:RETR}
\end{equation}
can be used. 
Note that the square root and inversion operation has lower priority than the $\diag$ operator 
(defined in notation section).
Geometrically, this retraction corresponds to 
the normalization to unit norm of each column of the matrix
$X+\xi_{X}$ to bring it back to $\mathcal{OB}(2,N)$ \cite{AMS08,AMSC14,MANOPT}. 

\section{Physical configurations of a ring laser cavity: The algorithm}

In the previous section, we have detailed how the gradient and Hessian 
of a function defined on $\mathcal{OB} (2,N)$ can be represented and 
used in solving the Newton equation \eqref{eq:NEM} for a generic function 
defined on $\mathcal{OB} (2,N)$.
This section shows how these results can be applied to 
obtain an algorithm to compute the physical configuration of 
a ring laser cavity with four spherical mirrors. 
From now on, therefore, $\mathcal{E}=\mathbb{R}^{3\times4}$ and $\mathcal{M}=\mathcal{OB}(2,4)$.

In the following, we detail how to compute the function $p:\mathcal{OB}(2,4) \rightarrow \mathbb{R}$ 
representing the laser path length. Recall that our goal is find the extremal points of this function 
as they represent the physical configurations consistent with Fermat's principle. 

\begin{figure}[htb]
\centering
\includegraphics[width=.9\textwidth]{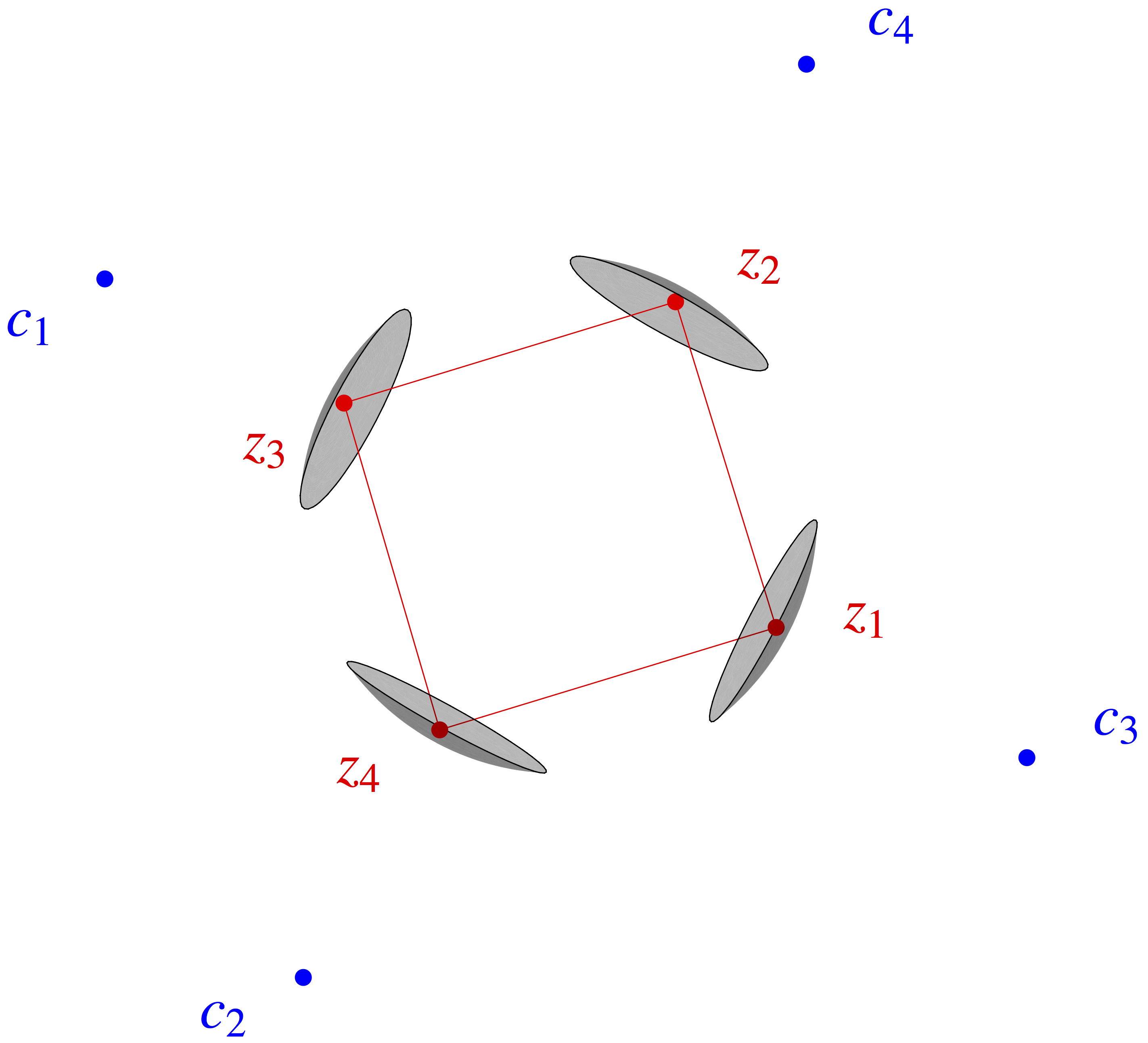}
\caption{\label{fig:CavityAndLaserBeam} An example of displacements of mirrors forming a square
optical cavity. Elements relative to each mirror are colored in grey.
Red dots mark the mirror centers, the red lines represent
the optical path length as a link between consecutive laser spots
$\mathbf{z}_{k}$.}
\end{figure}

The configuration of the optical cavity is 
\begin{equation}
X=\left(\mathbf{x}_{1},\dots,\mathbf{x}_{4}\right)\in\mathcal{OB}(2,4) , 
\end{equation}
while its parameters, namely the mirror centers and radii, are given, respectively, by
\begin{align}
  C & = \left(\mathbf{c}_{1},\dots,\mathbf{c}_{4}\right)\in\mathbb{R}^{3\times4}  , \\
  Q & = \text{diagonal}\left(r_{1},r_{2},r_{3},r_{4}\right)\in\mathbb{R}^{4\times4} .
\end{align}
In matrix form, the coordinates of the light spots on the mirror surface are given by
\begin{equation}\label{eq:spotposition4}
  Z(X) := XQ+C ,
\end{equation}
where $Z=(\mathbf{z}_{1},\mathbf{z}_{2},\mathbf{z}_{3},\mathbf{z}_{4}) \in \mathbb{R}^{3 \times 4}$. 
Figure~\ref{fig:CavityAndLaserBeam} provides a graphical representation of a typical 
disposition for the mirror centers, light spots, and laser beam path. 

The vectors joining consecutive light spots on the mirror surfaces, 
describing the sides of the polygonal cavity,
are given by $\mathbf{y}_{k}=(\mathbf{z}_{k+1}-\mathbf{z}_{k})$,
$k=\{1,2,3,4\}$, with the convention that $\mathbf{z}_{5}\coloneqq\mathbf{z}_{1}$. The
length of the optical path is therefore 
$ p(X;C,Q) = \sum_{i=1}^{4} || \mathbf{y}_{i} ||  $.
Defining $Y(X) := Z(X) M$ with 
\[
 M=
\begin{bmatrix} 
 \phantom{-}1 & \phantom{-}0 & \phantom{-}0 & -1 \\
-1                   & \phantom{-}1 &\phantom{-} 0 & \phantom{-}0  \\ 
 \phantom{-}0 & -1                  & \phantom{-}1 & \phantom{-}0 \\
 \phantom{-}0 & \phantom{-}0 & -1                  & \phantom{-}1
\end{bmatrix} ,
\]
the optical path length can be written, in matrix form, as
\begin{align}
  p(X;C,Q)= & \trace \left(\diag\left(Y^{T}(X)Y(X)\right)^{1/2}\right)\ . \label{eq:OPTPT4}
\end{align}
This is the formula that can be used to derive the Netwon method on $\mathcal{OB}(2,4)$.
Note that in \eqref{eq:OPTPT4},  the square root operator $(\cdot)^{1/2}$ acts 
component wise on the entries of the diagonal matrix and $\trace$ denotes the trace operator.

Equation~\eqref{eq:OPTPT4} also defines the extension $\bar{p}$  of $p$ to $\mathcal{E} = \mathbb{R}^{3 \times 4}$. 
The Euclidian gradient and Hessian of $ \bar{p} $, appearing in the formulas presented in Table~\ref{tab:CoordRepOB2N} and needed to computed the geometric gradient and Hessian of $p$,
are obtained from \eqref{eq:spotposition4} and \eqref{eq:OPTPT4} as
\begin{equation}
\left\langle \partial\bar{p}(X),\xi \right\rangle =\text{tr}\left[d\left(Y^{T}Y\right)^{-1/2}d\left(Y^{T}\xi_{X}QM\right)\right]\ \label{eq:4mg}
\end{equation}
and 
\begin{align}
\left\langle \partial^{2}\bar{p}(X)[\eta],\xi \right\rangle = & 
{\displaystyle \text{tr}\left[d\left(Y^{T}Y\right)^{-1/2}d\left(M^{T}Q \xi_{X}^{T}\eta_{X}QM\right)\right]} & \label{eq:4mh} 
\notag\\
& - {\displaystyle \text{tr}\left[d\left(Y^{T}Y\right)^{-3/2}d\left(Y^{T}\eta_{X}QM\right)d\left(Y^{T}\xi_{X}QM\right)\right]} .
\end{align}
Finally, the Riemannian gradient and Hessian of $p$ can be calculated from (\ref{eq:4mg}) and (\ref{eq:4mh}) 
recalling the general formulas (\ref{eq:RGM}) and (\ref{eq:RCM}) and the $\mathcal{OB}(2,N)$-specific 
expressions given in Table~\ref{tab:CoordRepOB2N}.
We have therefore all the ingredients to apply Algorithm~\ref{alg:NewtonMethodStatPoints}
described in Section~\ref{sec:StationaryPoints} to find the stationary point of $p$, i.e.,
the physical configurations of the laser path. 
En passant, we mention here that the vector area and compactness ratio of the cavity can be then computed 
simply as
$\mathbf{a}(X;C,Q)=\frac{1}{2}\sum_{k=1}^{N-1}\mathbf{z}_{k}\wedge\mathbf{z}_{k+1}$,
and 
$\mathbf{k}_{r}(X;C,Q)=\mathbf{a}(X;C,Q)/p(X;C,Q)$, respectively.

\section{Numerical Study}

We consider in this section a square optical cavity of side $L$ and four mirrors
with same radius $r$. Correspondingly, the cavity parameters are given by
\begin{equation}
C_{nom} = \left( r-\frac{L}{\sqrt{2}} \right) \left(\begin{array}{cccc}
1 & \:0 & -1 & \;0\\
0 & \:1 & \;0 & -1\\
0 & \:0 & \;0 & \;0
\end{array}\right)\;,
\end{equation}
and $ Q_{nom} = \mathrm{diagonal}(r,\, r,\, r,\, r) $.
The motivation for analyzing such configuration is the design
of control algorithms for the GINGER array of ring lasers \cite{GP5}.
In this situation, the optical cavity is made by spherical mirrors whose 
centers approximately lie on a planar square.

The physical configuration for this symmetry ring laser cavity can be computed by hand 
(and can be used as one of the tests to validate the correctness of the implementation of the algorithm presented 
in the previous section).  The physical configuration is 
\begin{align} 
  X^{*} & = -(r-L/\sqrt{2})^{-1}C_{nom} \in \mathcal{OB}(2,4)
  \label{eq:IdealSquareConf} 
\end{align}
and the corresponding laser path length is, by construction, $p(X^{*};C_{nom},Q_{nom})=4L$. 
It is interesting to compute the eigenvalues of $\mathrm{Hess}\, p$ at the point $X^{*}$.
Using the tools that we have created, one verifies that the spectrum of the geometric Hessian 
for this special configuration and parameters is 
\begin{align}
-\sqrt{2}&r\left(1,\,1,\,1-\frac{L}{\sqrt{2}r},\,1-\frac{L}{\sqrt{2}r},\,1-\frac{\sqrt{2}L}{r}\,,1-\frac{\sqrt{2}L}{r},\,1-\frac{\sqrt{2}L}{r},\,1-\frac{4L}{\sqrt{2}r}\right)
\label{eq:eigh}
\end{align}
All the eigenvalues of Hess$\, p$ are non-zero provided that the
ration $L/r$ is not equal to $\sqrt{2}$ , $1/\sqrt{2}$, or $\sqrt{2}/4$
which correspond to unstable optical cavity configurations \cite{GP4}.
For a stable optical cavity the Riemannian Hessian has non-zero eigenvalues, therefore $X^{*}$ is an isolated root of $\mathrm{grad}\, p$ and it is a \emph{saddle point} for the length function $p$, 
as the spectrum of the Hessian (\ref{eq:eigh}) has both strictly positive and negative eigenvalues.

Observe that, if a stable optical cavity is slightly misaligned from the nominal center configuration $C_{nom}$, 
by continuity arguments, one can conclude that the Riemannian gradient will keep having an isolated
root as the eigenvalues of the Riemannian Hessian will be different from zero
and that the Newton method is expected to converge.

The proposed geometric algorithm based on Newton's method 
the has been tested by Monte Carlo
techniques. Optical cavity configurations are generated starting with
mirror positions close to square configuration $C_{nom}$, with $L =
1.6$ m and represented by $\mathbb{R}^{ 3\times4 }$ random matrices whose entries
are uniformly distributed over the set $\left\{ \left|C_{ij}-C_{ij}^{*}\right|<\sigma\right\} $,
with $\sigma$ ranging from $10^{-6}L$ to $10^{-2}L$ with a logarithmic
spaced step of $L/10$. The radius matrix was set to $Q = r I_{4\times4}$,
with $r=4\,$ m. The chosen values correspond to the design of the GP2 ring
laser \cite{GP4}. The geometric algorithm has been applied
to find the saddle point of the function $p$, starting from the 
ideal square configuration given by \eqref{eq:IdealSquareConf}.
This procedure has been repeated $10^{4}$ times to assess whether mirror displacements are small enough for satisfying the convergence properties of the algorithm. 
All the execution runs showed no ill-conditioning problems in the Newton vector computation. 
In addition, the algorithm took {\em at most 3} iterations to generate a solution 
such that $\left\Vert \text{grad}\, p(x)\right\Vert <10^{-12}$ m, achieving quadratic convergence 
(expected as we are employing Netwon's method and the Hessian is not singular at the stationary point). 
In all the Monte Carlo runs, the computed laser spots positions are saddle points of $p$. 

\begin{figure}[htb]
\centering
\includegraphics[width=7cm]{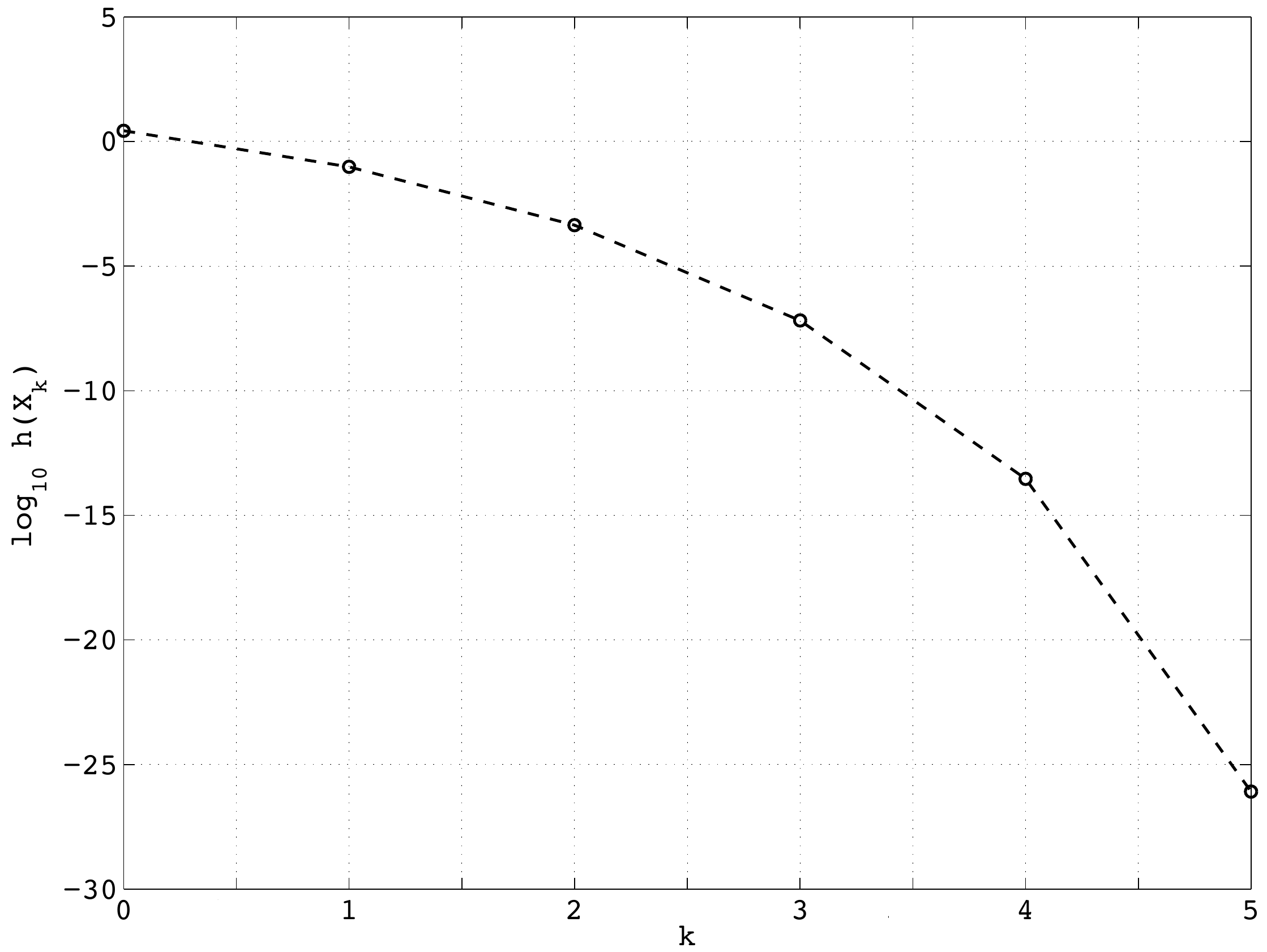}
\caption{Plot of $h(x_{k})=\left\Vert \nabla p(x_{k})\right\Vert $ versus the iteration index $k$. The data refers to a simulation run with $\sigma\sim0.5$ m. The algorithm converged in just $5$ iterations.}
\label{fig:1} 
\end{figure}

\begin{figure}[htb]
\centering
\begin{tabular}{cc}
\includegraphics[width=5.5cm]{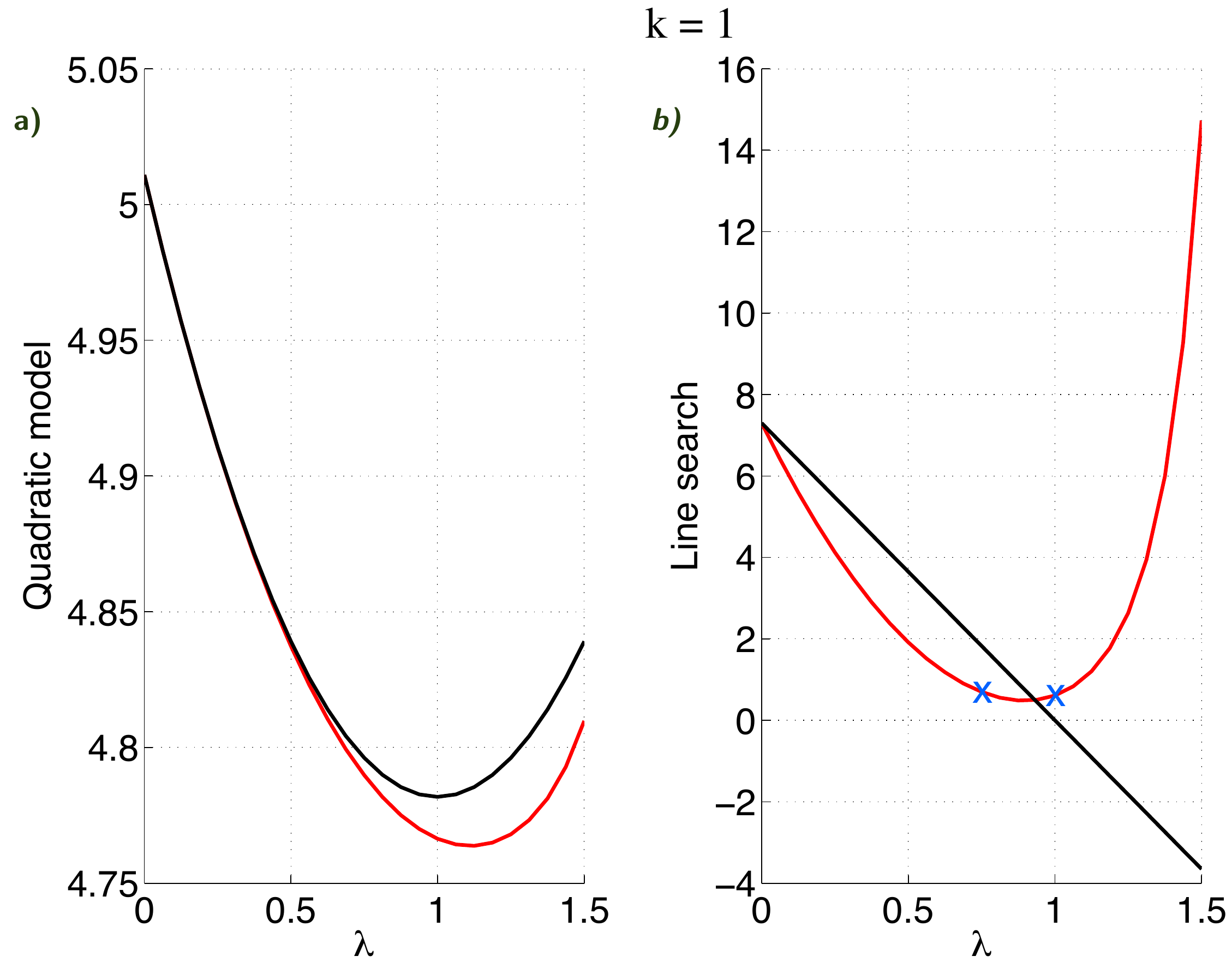} & \includegraphics[width=5.5cm]{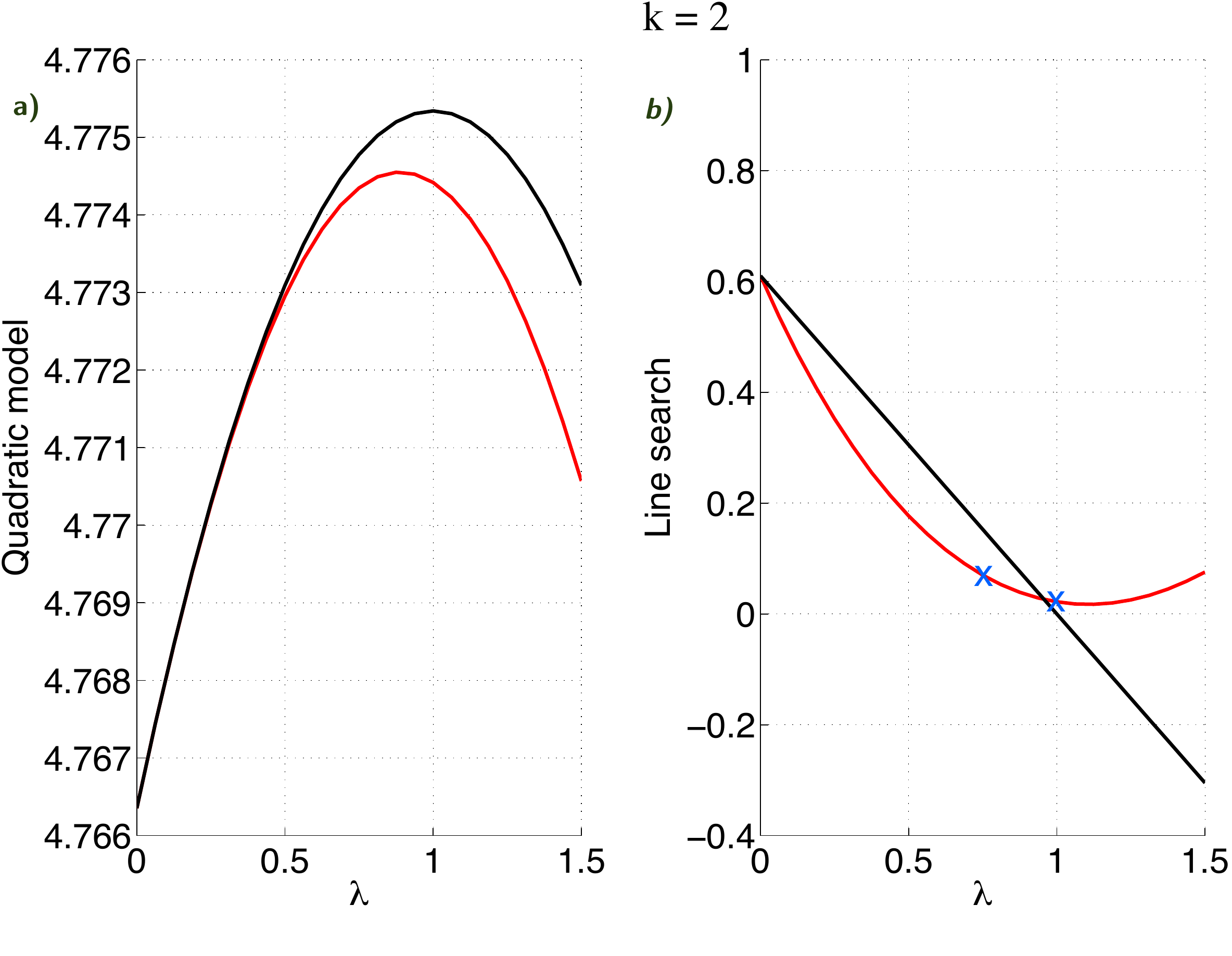}\tabularnewline
\includegraphics[width=5.5cm]{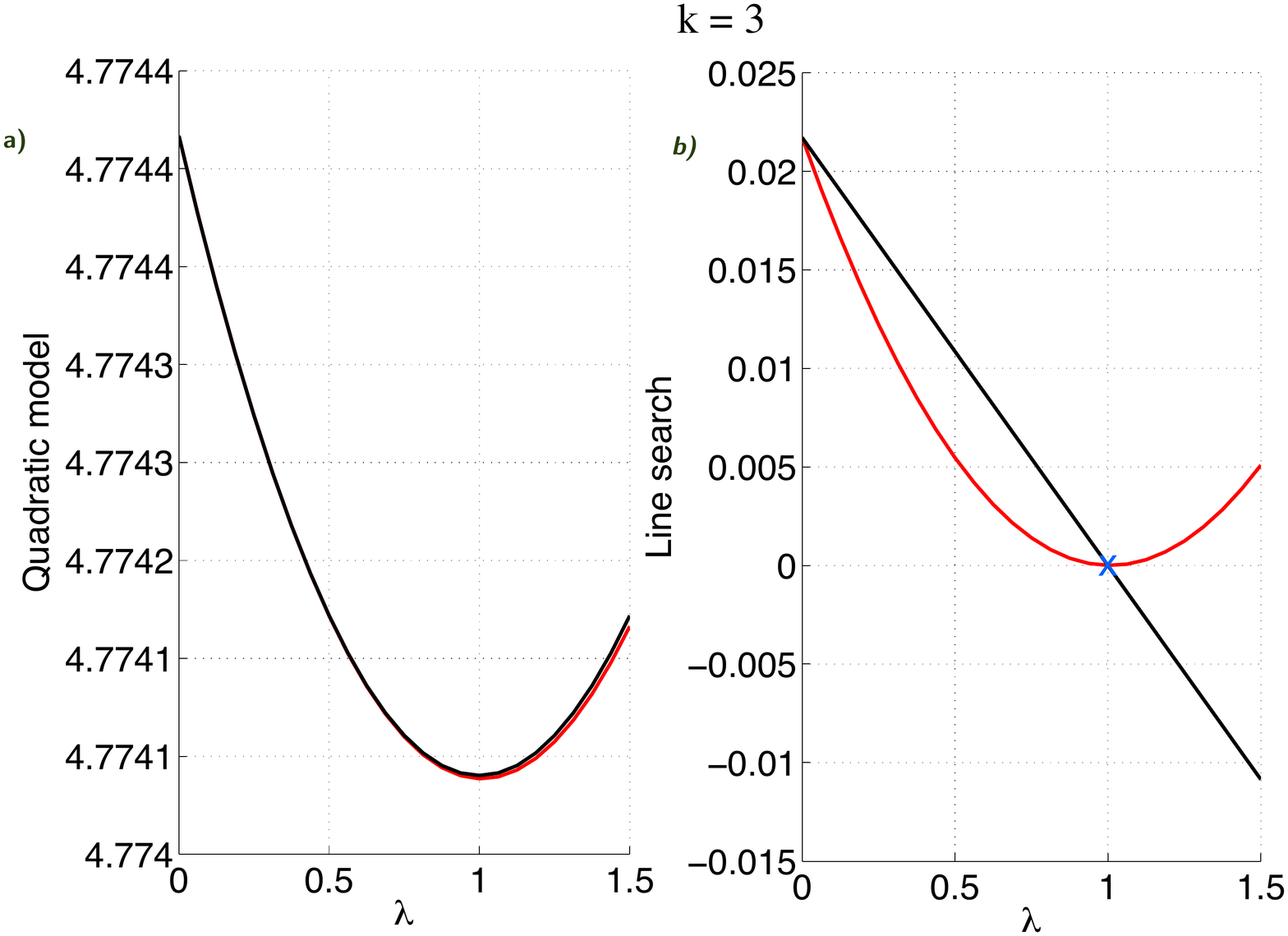} & \includegraphics[width=5.5cm]{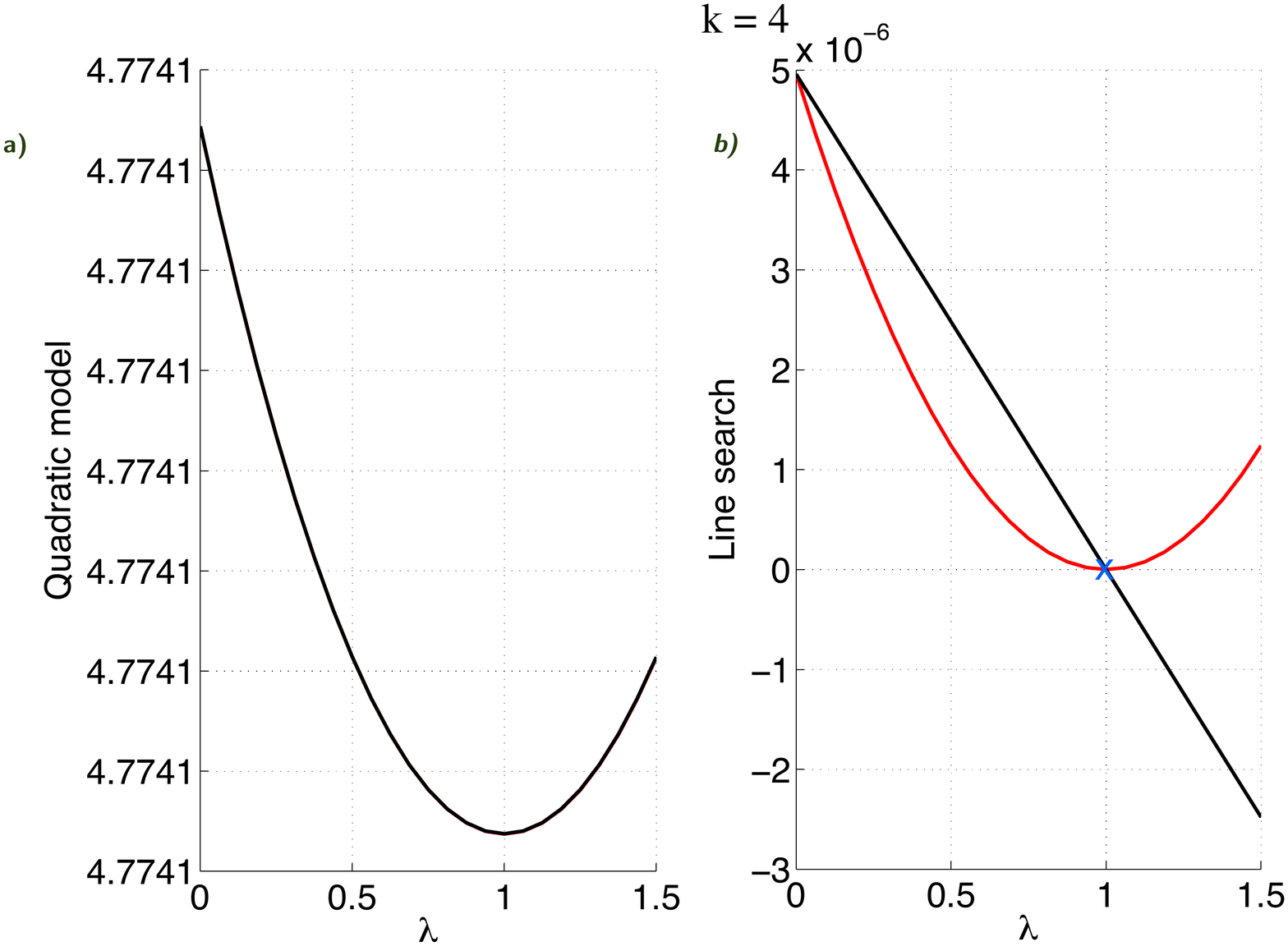}\tabularnewline
\includegraphics[width=5.5cm]{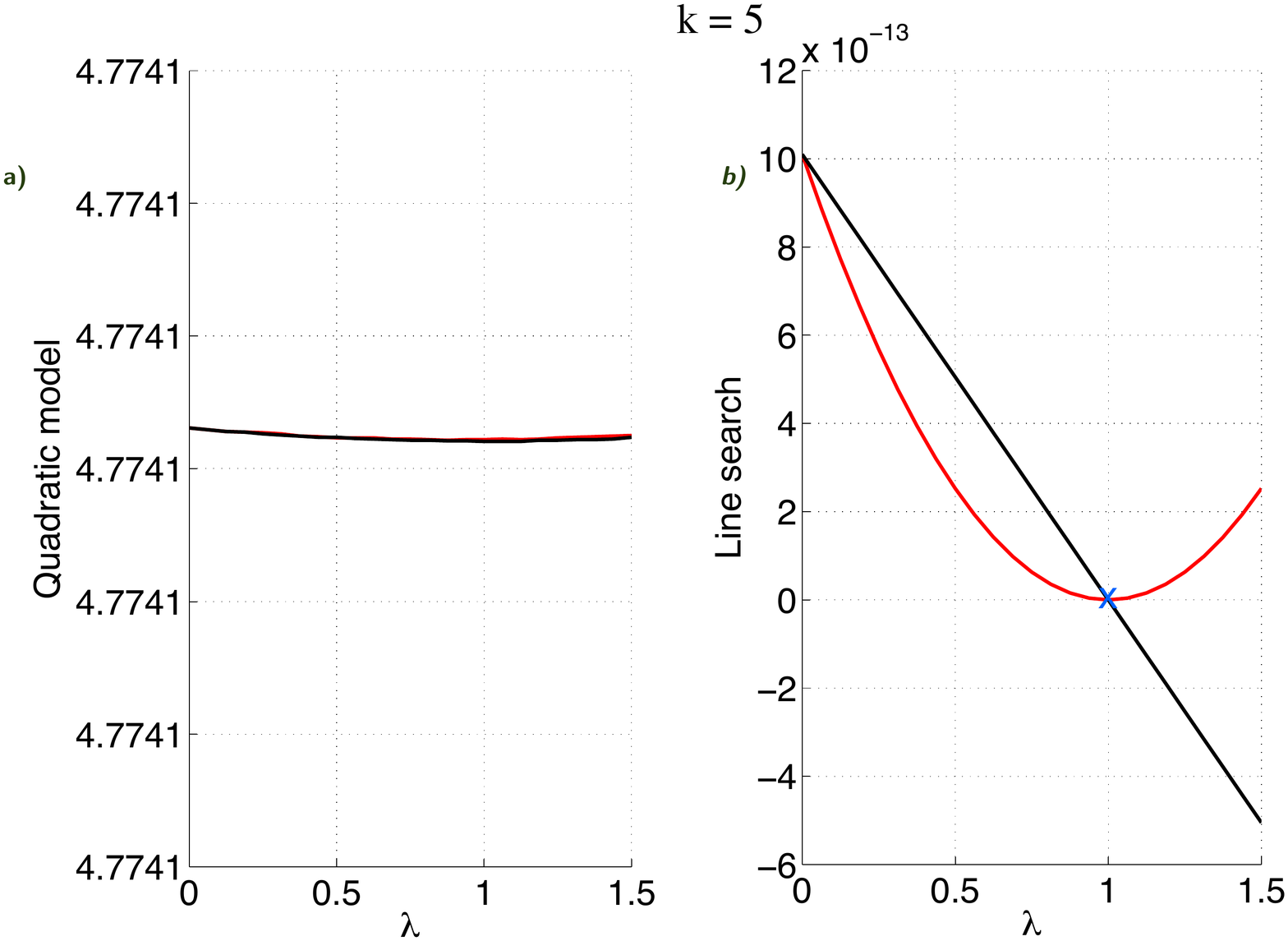} \tabularnewline
\end{tabular}
\caption{a): Comparison between the functions $p$ (red lines) and its second order geometric model $m_f$ (black lines). b): Comparison between the function $h$ (red lines) and $m_h$, related to the Armijo condition on $h$ (black lines). The blue crosses represent the iterates of the line search method. The plots are relative to consecutive algorithm iterations from 1 to 5.}
\label{fig:2}
\end{figure}

To better illustrate how the algorithm works, we show in Figure \ref{fig:1} the typical behavior of $\left\Vert \text{grad}\, p(x)\right\Vert $ in a run with $\sigma\sim0.5$ m, a value much larger than those considered in the previous simulations. The algorithm took $5$ iterations to converge. 
In Figure \ref{fig:2} two comparisons are displayed, between the function $p$ and its second order geometric model, and between the function $h$ and the Armijo condition for $h$, respectively.

We note that, at each Newton iteration, the geometric model better approximates $p$; in fact, from the third iteration the corresponding points in Figure \ref{fig:2} overlap. Therefore, at each Newton iteration, the Armijo threshold condition is fulfilled in a smaller number of line search iterations. As a consequence, the Armijo line search stops at its first iteration from the third Newton iteration. In the same figure, 
$m_{f}(\lambda)=f(x_{k})+\left\langle \text{grad}\, f(x_{k}),\lambda\eta_{x}\right\rangle +1/2\left\langle \text{Hess}\, f(x_{k})[\lambda\eta_{x}],\lambda\eta_{x}\right\rangle $, $m_{h}(\lambda)=h(x)+\sigma\left\langle \text{grad}\, h(x_{k}),\lambda\eta_{x}\right\rangle $, $\lambda\in[0,1]$, $\sigma=1/2$ and $\eta_{x}=\eta_{x_{k}}$ is
the Newton vector at the iteration $k$. 
In Figure~\ref{fig:2}, each of the 5 iterations of the run reported in Figure \ref{fig:1} is displayed 
Note how the optical path length $p$ is well modeled by the second order approximation 
$m_{f}(\lambda)$ almost from the first iteration and this allows for step sizes $t_{k}$ to be close to one 
from the beginning as in a pure Newton method without the line search condition (\ref{eq:CARM}).
Clearly, this is due also by the fact that we provide the algorithm with a good initial condition.

\section{Conclusions}

We have addressed the problem of computing the light path in optical cavities as a function of the mirror positions and orientations by means of the Fermat's principle. To find the stationary optical path in a polygonal cavity, a geometric algorithm based on Newton method is proposed, exploiting the embedding of the Oblique Manifold $\mathcal{OB}(2,N)$ in the Euclidean space $\mathbb{R}^{3\times N}$ for the computation of Riemannian gradient and Hessian of the optical path length. The Riemannian gradient and Hessian are then used in each iteration of the algorithm. The algorithm  exhibits second order convergence rate.

Our approach is motivated by the need to compute the optical path of the laser beams in resonant cavities and the algorithm has the potential to allow for the application of control techniques to constrain the beams path.

We showed that the distance function meets the regularity requirements
in a neighborhood of the solution for the perfectly aligned square cavity. 
Monte Carlo simulations have shown that this neighborhood contains all the cavity configurations that can be encountered in practice for heterolithic square ring lasers. In fact, our algorithm has been reported to evaluate the beam steering of a square cavity even with mirror positioning errors $\left\Vert C-C^{*}\right\Vert 10^{-3}$ m, while precision machinery of ring laser frames ensure mirrors positioning within $\sim10^{-5}$ m. 

The proposed geometric algorithm provides a relative accuracy of $1$ part in $10^{16}$ in evaluating the optical cavity configuration of a square ring laser. It is worth noticing that greater precisions can be achieved, even if they are not of physical interest. The computational cost of the proposed method is very low, since at most 3 iterations are required to reach the desired accuracy in the Riemannian gradient norm in Monte Carlo simulations. 

The geometric approach described in the present paper seems to be well suited to deal with geometrical optics problems where Gaussian profiles are used for beam description. Based on the presented algorithm, future work will be devoted to the calibration and active control of resonant optical cavities \cite{Future}.

\appendix
\section{Discussion and Convergence of Algorithm \ref{alg:GNA}}
In Algorithm \ref{alg:GNA} at each iteration the Newton equation
(\ref{eq:NEM}) is solved for the function $f$, then the function
$h(x)=\left\Vert \text{grad}\, f(x)\right\Vert ^{2}$ is minimized
along the computed direction. In this way we need to compute only
Hess$\, f$ and grad$\, f$, avoiding the computation of Hess$\, h$,
that would require to compute the third derivative of $f$. 
\begin{proposition}
Algorithm \ref{alg:GNA} converges to the stationary point $x^{*}$
of the function $f$ with quadratic convergence rate, provided that,
in a neighborhood $\mathcal{I}(x^{*})$ of $x^*$, $\text{grad}\, f\neq0$,
$\text{Hess}\, f$ is injective, and the first iterate is $x_{0}\in\mathcal{I}(x^{*})$. \end{proposition}
{\it Proof } Let $x$ denote a generic algorithm iterate, By hypotheses
the Newton vector $\eta_{x}$, solution of (\ref{eq:NEM}), is well
defined. The Riemannian gradient of $h$ reads

\begin{equation}
\text{grad}\, h=2\text{Hess}\, f[\text{grad}\, f]\ .\label{eq:gradh}
\end{equation}
By evaluating the expression $Dh(x)[\eta_{x}]$ we get

\begin{align}
Dh(x)[\eta_{x}]= & 2\left\langle \text{grad}\, f(x),\text{Hess}\, f(x)[\text{Hess}\, f(x)^{-1}[-\text{grad}\, f(x)]]\right\rangle \label{eq:DERHETA}\\
= & -2\left\Vert \text{grad}\, f(x)\right\Vert ^{2}\nonumber \\
= & -2h(x).\nonumber 
\end{align}
The sequence $\left\{ \eta_{x_{k}}\right\} $ is gradient related
to $\left\{ x_{k}\right\} $. In fact by hypothesis and (\ref{eq:gradh})
it holds grad$\, h(x_{k})\neq0$, therefore, using (\ref{eq:DERHETA})
we get $-2\sup_{\mathcal{I}(x^{*})}h(x_{k})=\sup_{\mathcal{I}(x^{*})}Dh(x_{k})[\eta_{x_{k}}]<0$.
By the smoothness of the functional Hess$\, f$ and of the vector
field grad$\, f$, since $\mathcal{I}(x^{*})$ is a compact set, we
can conclude that $\left\{ \eta_{x_{k}}\right\} $ is bounded. Hence
Algorithm \ref{alg:GNA} fits in the framework of Theorem 4.3.1 and
Theorem 6.3.2 \cite[Chs.4-6]{AMS08}, stating that every accumulation
point of $\left\{ x_{k}\right\} $ is a critical point of $h$, so
that the local quadratic convergence holds.
\qed

Note that the Armijo condition (\ref{eq:CARM}) for the function $h$
and the direction $\eta_{x}$ can be rewritten as

\begin{align}
h(x)-h\left(y_{k}\right)<-\sigma\gamma_{k}Dh(x)[\eta_{x}] & =2\sigma\gamma_{k}h(x)\label{eq:ARMCR}\\
h\left(y_{k}\right)>\left(1-2\sigma\gamma_{k}\right)h(x) & \ ,
\end{align}

where $y_{k}=R_{x}(\gamma_{k}\eta_{x})$, $x=x_{k}$, $\eta_{x}=\eta_{x_{k}}$,
and $k$ is the iteration number.

\end{document}